\documentclass[aps,preprint,showpacs,floatfix]{revtex4-1}
\usepackage{bm}
\usepackage{amsmath}
\usepackage{epsfig}
\usepackage{rotating}
\usepackage{graphicx}

\def\pr{\prime}
\def\be{\begin{equation}}

\def\ee{\end{equation}}
\def\barr{\begin{array}}
\def\earr{\end{array}}

\def\l{\left}
\def\r{\right}
\def\dis{\displaystyle}
\def\ed{\end{document}}

\def\can{{\cal N}}

\def\pr{\prime}

\def\la{\lambda}
\def\ed{\end{document}}
\begin{document}

\title{Next highest weight and other lower $SU(3)$ irreducible representations
with proxy-$SU(4)$ symmetry for nuclei with $32 \le \mbox{Z,N} \le 46$}

\author{V.K.B. Kota\footnote{
{\it E-mail address:}
vkbkota@prl.res.in (V.K.B. Kota)}}

\affiliation{Physical Research  Laboratory, Ahmedabad 380 009, India}

\begin{abstract}

In the applications of proxy-SU(3) model in the context of determining $(\beta,\gamma)$ values for nuclei across the periodic table, for understanding the preponderance of triaxial shapes in nuclei with $Z \ge 30$, it is seen that one needs not only the highest weight (hw) or leading $SU(3)$ irreducible representation (irrep) $(\lambda_H, \mu_H)$ but also the lower $SU(3)$ irreps $(\lambda ,\mu)$ such that $2\lambda + \mu =2\lambda_H + \mu_H-3r$ with $r=0,1$ and $2$ [Bonatsos et al., Symmetry {\bf 16}, 1625 (2024)]. These give the next highest weight (nhw) irrep, next-to-next highest irrep (nnhw) and so on. Recently, it is shown that for nuclei with $32 \le \mbox{Z,N} \le 46$, there will be not only proxy-$SU(3)$ but also proxy-$SU(4)$ symmetry [Kota and Sahu, Physica Scripta {\bf 99}, 065306 (2024)]. Then one has the algebra $U(10n) \supset \l[U(10 \supset SU(3) \supset SO(3)\r] \otimes SU(n)$; $n=2$ when there are only valence protons or neutrons and $n=4$ for nucleons with isospin $T$ [with $n=4$ we have proxy-$SU(4)$ symmetry]. Following these developments, presented in this paper are the $SU(3)$ irreps $(\lambda ,\mu)$ with $2\lambda + \mu =2\lambda_H + \mu_H-3r$, $r=0,1,2$ for various isotopes of Ge, Se, Kr, Sr, Zr, Mo, Ru and Pd assuming good proxy-$SU(4)$ symmetry. A simple method for obtaining the SU(3) irreps is described and as a test, results for identical nucleons are used. The tabulations for proxy-$SU(3)$ irreps provided in this paper, for Ge, Se, Kr, Sr, Zr, Mo, Ru and Pd isotopes with $32 \le \mbox{N} \le 46$ and good proxy-$SU(4)$ symmetry, will be useful in further investigation of triaxial shapes in these nuclei. 

\end{abstract}

%\pacs{}

\maketitle

\section{Introduction}

Elliott showed for the first time in 1958 that the oscillator orbital $SU(3)$ symmetry generates rotational spectra within the spherical shell model of atomic nuclei \cite{Ell-58a,Ell-58b}. Implicit here also is the goodness of Wigner's spin-isospin $SU(4)$ symmetry \cite{Wig,JCP}. Overcoming the breaking of $SU(3)$ symmetry due to the strong shell model spin-orbit force, $SU(3)$ model appeared in many different forms in nuclear structure. Some of these are: pseudo-$SU(3)$ model based on pseudo spin and pseudo Nilsson orbits \cite{JPD1,JPD2}, $SU(3)$ in the $Sp(6,R)$ model \cite{Rowe-1,Rowe-2}, $SU(3)$ with the interacting boson, boson-fermion and boson-fermion-fermion models \cite{Iac-87,Iac-91,BK,UK}, $SU(3)$ in various cluster models \cite{Ch-1}, and so on; see \cite{Ko-book} for more details and applications. Most recent addition to all these is the proxy-$SU(3)$ model introduced by Bonatsos et al \cite{Bona-1,Bona-2,Bona-3,Bona-4} and  the present article belongs to this model.

Proxy-$SU(3)$ model was introduced in 2017 for heavy deformed nuclei with protons and neutrons occupying different shells \cite{Bona-1}. This scheme is defined through the replacement of intruder orbitals in a given shell by orbitals dropped into the lower shell of the same type of nucleons (protons or neutrons). Then, the intruder high-$(\eta, \ell, j)$ orbit that is pushed down due to strong spin-orbit force is replaced by the proxy $(\eta-1,\ell-1,j-1)$ orbit by ignoring the high-lying $(\eta,\ell,j:k=\pm j)$ states. Then, for example $^0h_{11/2}$ in 50-82 shell changes to proxy $^0g_{9/2}$ and the 50-82 shell becomes proxy $\eta=4$ shell. Similarly, the 82-126 shell becomes proxy $\eta=5$ shell and so on. With the proxy oscillator shells, we have proxy-$SU(3)$ symmetry in each shell and by coupling these proton $SU(3)$ and neutron $SU(3)$ algebras, we have proxy-$SU(3)$ symmetry for heavy nuclei.
With the highest weight (hw) or the leading $SU(3)$ irreducible representation (irrep), labeled in Elliott's notation by
$(\lambda_H, \mu_H)$, describing ground state structure, the hw $SU(3)$ irrep given by the proxy-$SU(3)$ scheme is found to describe prolate shape dominance over oblate shape in nuclei with protons ($p$) and neutrons ($n$) occupying different oscillator
shells and in some situations the same shell \cite{Bona-2,Bona-3}. This is the first significant achievement of the proxy-$SU(3)$ model \cite{Bona-1,Bona-2,Bona-3,Bona-4}. Bonatsos et al in their studies generated the hw $SU(3)$ irreps for all particle numbers in oscillator shells with shell number $\eta = 2$, 3, 4, 5 and 6 using the computer codes due to Draayer et al \cite{jpd-c1,jpd-c2} and then the stretched coupling of the hw proton and hw neutron $SU(3)$ irreps gives the hw $SU(3)$ irrep for the total pn-system. However, later a simple formula for the hw $SU(3)$ irreps was presented in \cite{Ko-hw} (see also Section II).

Going further, more recently the proxy$SU(3)$ model is applied in the study of shape coexistence and preponderance of triaxial shapes in nuclei across the periodic table \cite{Bona-5,Bona-6,Bona-7,Bona-8}. Crucial element in these analysis is the calculation of the deformation parameters $(\beta , \gamma)$
that correspond to a given hw $SU(3)$ irrep $(\lambda_H, \mu_H)$. This is facilitated by the simple formulas derived by Draayer et al \cite{jpd-bg} for $\beta$ and $\gamma$ in terms of $\lambda$ and $\mu$ values of any $(\la ,\mu)$ irrep. In a detailed study of triaxial shapes (see \cite{Otsuka} for the recent recognition that triaxial shapes are much more common in nuclei than what was thought about in the past), it is seen that the next-highest-weight (nhw) proxy-$SU(3)$ irrep will play an important role in better determination of the triaxiality parameter $\gamma$. Following these, Bonatsos et al determined (again using the codes in \cite{jpd-c1,jpd-c2}) nhw $SU(3)$ irreps for nuclei all across the periodic table and tabulated the same \cite{Bona-5,Bona-7}. 
In all these, the proton-neutron proxy-$SU(3)$ formulation is used, along with some other assumptions, as most nuclei considered have valence protons and neutrons in different proxy shells. Further, the authors presented the nhw $SU(3)$ irreps also for nuclei with $ 36 \le \mbox{Z,N} \le 46$ \cite{Bona-5}. However, recently in \cite{Ko-psu4} it was pointed out that proxy-$SU(4)$ symmetry is important for nuclei with $32 \le \mbox{Z,N} \le 46$ (Ge to Pd isotopes); in these nuclei the valence protons and neutrons occupy the same proxy oscillator shell with shell number $\eta=3$. Thus, for these nuclei determination of nhw and next-to-next highest weight (nnhw) and other lower proxy-$SU(3)$ irreps assuming proxy-$SU(4)$ symmetry becomes important. Giving results with proxy-$SU(4)$ symmetry, for Ge to Pd isotopes, for the lower $SU(3)$ irreps is the purpose of the present article. 

Before going further, it is important to mention that with spin-isospin proxy-$SU(4)$ symmetry for Ge to Pd isotopes, there will be not only the orbital proxy-$SU(3)$ symmetry but also proxy-$SU(5)$, $SU(4)$ and $SO(10)$ symmetries \cite{Ko-psu4}. It is possible that these also may play an important role in the study of triaxiality and shape coexistence in Ge to Pd isotopes. This will be investigated in future elsewhere. Now we will give a preview.

In Section II described is a simple method adopted in the present work for obtaining $SU(3)$ irreps for a given number of nucleons in a oscillator shell $\eta$ with a given isospin value. As a test of the method, results for identical nucleons are presented
and compared with those given in \cite{Bona-5,Bona-7}. Section III contains the main results for the proxy-$SU(3)$ irreps for Ge to Pd isotopes where proxy-$SU(4)$ symmetry is used. Present results are compared with those given recently by Bonatsos et al \cite{Bona-5} and pointed out the differences.
Finally, Section IV gives conclusions.

\section{A simple method for obtaining $SU(3)$ irreps and results for two column irreps of $U(\can)$}

\subsection{Method for two and four column $U(\can)$ irreps reductions}

With nucleons in a oscillator shell $\eta$, we have $SU(3)$ algebra with $U(r\can) \supset [U(\can) \supset SU(3) \supset SO(3)] \otimes SU(r)$; $\can = (\eta+1)(\eta+2)/2$. Note that  $r=2$ for identical nucleons and 4 for nucleons with isospin. 
As we have direct product algebra $U(r\can) \supset U(\can) \otimes SU(r)$, the $SU(r)$ irreps $\{F\}$ uniquely define the $U(\can)$
irreps $\{f\}$. For $r=2$ the $SU(r)$ irreps $\{F\}$ will be two rowed $\{F\}=\{F_1,F_2\}$ giving spin $S=F_1-F_2$ [in fact $\{F\}$ is a $U(2)$ irrep]. For $m$ number of identical nucleons $m=F_1+F_2$ with $F_1 \ge F_2 \ge 0$. With this, the corresponding $U(\can)$ irreps will be of the form $\{2^a\;1^b\}$ satisfying $2a+b=m$ and $S=b/2$. Therefore, for spin $S=0$ systems the $U(\can)$ irreps will be of the form $\{2^n\}$ with $n=m/2$ (note that $m$ must be even for $S=0$). Going to $r=4$, we have the spin-isospin $SU(4)$ algebra. Then, the $U(4)$ irreps $\{F\}$ will be maximum four rowed, $\{F\} = \l\{F_1,F_2,F_3,F_4\r\}$ with $F_1 \ge F_2 \ge F_3 \ge F_4 \ge 0$ and $m=F_1+F_2+F_3+F_4$. This gives the corresponding $U(\can)$ irrep $\{f\}$, conjugate to $\{F\}$, to be $\{f\}=\{4^a 3^b 2^c 1^d\}$ with $a=F_4$, $b=F_3-F_4$, $c=F_2-F_3$ and $d=F_1-F_2$; note that
$\{f\}$ is maximum four columned.

With valence protons and neutrons in the same oscillator shell $\eta$, we assume that the Wigner's spin-isospin $SU(4)$ symmetry is good. Now, given the nucleon number $m$ and the isospin $T=|T_Z|$ (note that $T_Z=(m_p-m_n)/2$ where $m_p$ is number of valence protons and $m_n$ is number of valence neutrons with $m=m_p+m_n$), the lowest $U(4)$ irrep $\{F_1,F_2,F_3,F_4\}$ is 
given as follows (note that $SU(4)$ irreps follow from $U(4)$ irreps) \cite{JCP,MK-su4,Piet-su4}. In this paper we will restrict to the situation with $m$ always even. Then we have,
\be
\barr{l}
\l\{F_1,F_2,F_3,F_4\r\} = \l\{\frac{m+2T}{4}, \frac{m+2T}{4}, \frac{m-2T}{4}, 
\frac{m-2T}{4}\r\}\;\;\;\mbox{for}\;\;\;\frac{m}{2}+T\;\;\mbox{ even}\;,\\
\l\{F_1,F_2,F_3,F_4\r\}=\l\{\frac{m+2T+2}{4}, \frac{m+2T-2}{4}, 
\frac{m-2T+2}{4}, \frac{m-2T-2}{4}\r\}\;\;\;\mbox{for}\;\;\;
\frac{m}{2}+T\;\;\mbox{odd}\;.\\
\earr \label{su4-1}
\ee
The only exception is $T=0$ for $m=4r+2$ type and then 
\be
\l\{F_1,F_2,F_3,F_4
\r\}=\l\{\frac{m+2}{4}, \frac{m+2}{4}, \frac{m-2}{4}, \frac{m-2}{4}\r\}\;.
\label{su4-2}
\ee
Using  Eqs. (\ref{su4-1}) and (\ref{su4-2}) it is easy to obtain the lowest $U(4)$ irrep for a given $m$  and $T=|T_z|$, i.e. for a given even-even nucleus. Our purpose here is to develop a simple method (that can be converted into a computer code) for obtaining $\{f\}_{U(\can)} \rightarrow (\la,\mu)_{SU(3)}$ reductions with $\{f\}$ maximum of four columns.

In the first step, all possible antisymmetric irreps $\{1^r\}$ of $U(\can)$, with $r \le \can/2$, are reduced into $SU(3)$ irreps using the difference method described in Section 3.2.3 of \cite{Ko-book}. In fact tabulations for $\eta=3$ ($\can=10$),
$\eta=4$ ($\can=15$), $\eta=5$ ($\can=21$) are given in \cite{Ko-book}. Complete tabulations for $\eta=6$ ($\can=28$) can be obtained from the author. For $\{1^r\}$ with $r > \can/2$, the reductions follow from the rule,
\be
\barr{l}
\l\{1^r\r\} \rightarrow \dis\sum_i x_i (\la_i , \mu_i) \oplus \\
\Rightarrow \l\{1^{\can -r}\r\} \rightarrow \dis\sum_i x_i (\mu_i , \la_i) \oplus\;.
\earr \label{su4-3}
\ee
Here, $x_i$ is the multiplicity of the irrep $(\la_i , \mu_i)$ implying that the irrep $(\la_i , \mu_i)$ appears $x_i$ times [$x_i=0$ implies that the irrep $(\la_i , \mu_i)$ will not appear in the reduction]. It is easy to enumerate all $(\la_i , \mu_i)$,
for a given $r$, by enumerating the $U(3)$ irreps $(n_1,n_2,n_3)$ where $n_1 \ge n_2 \ge n_3 \ge 0$ with $n_1+n_2+n_3=\eta r$. Then, $\la = n_1-n_2$ and $\mu = n_2-n_3$. Note that with the reductions for $\{1^r\}$ of $U(\can)$ are all available, we will
generate the reductions for any 4 column irrep $\{f\}$ of $U(\can)$ by
expanding $\{f\}$ as a $4 \times 4$ determinant involving only
antisymmetric irreps of $U(\can)$. Given $\{f\}=\{4^a 3^b 2^c 1^d\}$ irrep, we have \cite{Wy-70,Little},
\be
\barr{l}
\l\{4^a 3^b 2^c 1^d\r\} = \l|\barr{cccc} \l\{1^{F_1}\r\} &  
\l\{1^{F_1+1}\r\} &  \l\{1^{F_1+2}\r\} & \l\{1^{F_1+3}\r\} \\
\l\{1^{F_2-1}\r\} &  \l\{1^{F_2}\r\} &  \l\{1^{F_2+1}\r\} & \l\{1^{F_2+2}\r\} \\ \l\{1^{F_3-2}\r\} &  \l\{1^{F_3-1}\r\} &  \l\{1^{F_3}\r\} & \l\{1^{F_3+1}\r\} \\ \l\{1^{F_4-3}\r\} &  
\l\{1^{F_4-2}\r\} &  \l\{1^{F_4-1}\r\} & \l\{1^{F_4}\r\} \earr \r|\;;\\
F_1=a+b+c+d,\;\;F_2=a+b+c,\;\;F_3=a+b,\;\;F_4=a\;.
\earr \label{su4-4}
\ee
The multiplications in the determinant in Eq. (\ref{su4-4}) are Kronecker multiplications. Also, $\{1^x\}=0$ if $x < 0$ and $\{1^x\}=\{0\}$ if $x=0$. It important to note that when we expand the determinant Eq. (\ref{su4-4}), there will be 24 terms of the type 
$$
\l\{1^{X_1}\r\} \l\{1^{X_2}\r\} \l\{1^{X_3}\r\} \l\{1^{X_4}\r\}
$$
where $X_1+X_2+X_3+X_4=m$ if $\{f\}$ is an irrep of a $m$ nucleon system. The product of the $\{1^{X_i}\}$ is reduced to $SU(3)$ irreps as follows. Let us first consider the reduction of 
$\l\{1^{X_1}\r\} \l\{1^{X_2}\r\}$. To this end we start with
(see Eq. (\ref{su4-3})),
\be
\barr{l}
\l\{1^{X_1}\r\} = \dis\sum_i x_i (\la_{1i} , \mu_{1i})\;, \\
\l\{1^{X_2}\r\} = \dis\sum_j y_j (\la_{2j} , \mu_{2j})\;. 
\earr \label{su4-5}
\ee
and then we need
\be
(\la_{1i} , \mu_{1i}) \times (\la_{2j} , \mu_{2j}) \rightarrow
\dis\sum_k \kappa^{(ij)}_{12:k} (\la_{12:k} , \mu_{12:k})\;.
\label{su4-6}
\ee
The $(\la_{12:k} , \mu_{12:k})$ here are all the $SU((3)$ irreps for $X_{12}=X_1+X_2$ number of particles and they are enumerated using the procedure described just after Eq. (\ref{su4-3}). The multiplicity coefficients $\kappa^{(ij)}_{12:k}$ follow easily from the rule given by Chew and Sharp \cite{CS}. This is converted into a small programme called MULTU in the Akiyama and Draayer $SU(3)$ package \cite{jpdaky}. Then we have,
\be
\barr{l}
\l\{1^{X_1}\r\} \l\{1^{X_2}\r\} = \dis\sum _k z_k (\la_{12:k} , \mu_{12:k})\;;\\
z_k = \dis\sum_{i,j} x_i y_j \kappa^{ij}_{12:k}\;.
\earr \label{su4-7}
\ee
Starting with Eq. (\ref{su4-7}) and $\l\{1^{X_3}\r\}$ decomposition to $SU(3)$ irreps $(\la_{3\ell} , \mu_{3\ell})$
with multiplicities say $p_{3\ell}$, and repeating the procedure that gave Eq. (\ref{su4-7}) but with the $SU(3)$ irreps $(\la_{123:l} , \mu_{123:l})$ for $X_{123}=X_1+X_2+X_3$ number of particles, we will obtain
$$
\l\{1^{X_1}\r\} \l\{1^{X_2}\r\} \l\{1^{X_3}\r\}
$$
reduction to $SU(3)$ irreps. Repeating the same procedure we will finally obtain the reduction of 
$$
\l\{1^{X_1}\r\} \l\{1^{X_2}\r\} \l\{1^{X_3}\r\} \l\{1^{X_4}\r\}
$$
to $SU(3)$ irreps with the associated multiplicities. Applying this to the 24 terms in the determinant in Eq. (\ref{su4-4}) and adding them we will have the final result for 4-column $\{f\}$ reduction to $SU(3)$ irreps. Note that two column irreps involve only $2 \times 2$ matrices and hence their reduction are obtained more easily by just using Eqs. (\ref{su4-5})-(\ref{su4-7}). Using the above procedure, we have developed a simple FORTRAN code for obtaining the reduction of any two or four column irrep $\{f\}$ of $U(\can)$ to $SU(3)$ irreps. The final results for any $\l\{f\r\}$ of $U(\can)$ are of the form,
\be
\l\{f\r\} = \dis\sum_q \Gamma_q \l(\la_q , \mu_q\r) \oplus \;.
\label{su4-9}
\ee
where $\Gamma_q$ is the multiplicity of the $SU(3)$ irrep $\l(\la_q , \mu_q\r)$ and the irreps $\l(\la_q , \mu_q\r)$ are
all the $SU(3)$ irreps of $m=\sum_{i=1}^{\can}\; f_i$. A good check of the reductions obtained is the dimensionality check.
A simple formula for $U(\can)$ irreps dimension and $SU(3)$ irreps dimension along with Eq. (\ref{su4-9}) will give \cite{Wy-70,Little},
\be
d(\l\{f\r\}) = \dis\prod_{i<j=1}^{\can} \dis\frac{f_i-f_j+j-i}{j-i} = \dis\sum_q \Gamma_q \;(\la_q+1)(\mu_q+1)(\la_q +\mu_q+2)/2\;.
\label{su4-10}
\ee

\begin{table}
\caption{Results for $U(10) \supset SU(3)$ reductions with $\eta=3$ for identical fermions with spin $S=0$, i.e. for all $\{2^r\}$ irreps of $U(10)$ with $r=0$ to $10$. The 'd' in the table gives dimension of the irrep $\l\{2^r\r\}$. In the table,$\Gamma$  in $\Gamma(\la\;\;\mu)$ is the multiplicity of the $SU(3)$ irrep $(\la\;\;\mu)$. The irreps with $\Gamma=0$ are not shown as they will not exist in the reductions.}
{\scriptsize{
\begin{tabular}{l}
\hline
$\l\{0\r\}\;;\;\;d=1$ \\
\verb|   1( 0  0) | \\
$\l\{2\r\}\;;\;\;d=55$ \\
\verb|   1( 6  0)     1( 2  2) | \\
$\l\{2^2\r\}\;;\;\;d=825$ \\
\verb|  1( 8  2)    1( 7  1)    2( 4  4)    1( 5  2)    1( 6  0)    1( 3  3)    1( 4  1)    1( 0  6)   | \\
\verb|  1( 1  4)    2( 2  2)    1( 1  1) | \\
$\l\{2^3\r\}\;;\;\;d=4950$ \\
\verb|  1(12  0)    1( 9  3)    1( 6  6)    1( 7  4)    3( 8  2)    2( 5  5)    3( 6  3)    2( 7  1)   | \\
\verb|  1( 2  8)    2( 3  6)    5( 4  4)    3( 5  2)    4( 6  0)    1( 1  7)    2( 2  5)    5( 3  3)   | \\
\verb|  3( 4  1)    3( 0  6)    2( 1  4)    5( 2  2)    1( 3  0)    1( 0  3)    1( 1  1)    2( 0  0)   | \\
$\l\{2^4\r\}\;;\;\;d=13860$ \\
\verb|  1(10  4)    1(12  0)    1( 8  5)    2( 9  3)    1(10  1)    1( 5  8)    3( 6  6)    3( 7  4)   | \\
\verb|  5( 8  2)    1( 9  0)    1( 3  9)    2( 4  7)    5( 5  5)    7( 6  3)    5( 7  1)    1( 0 12)   | \\
\verb|  4( 2  8)    6( 3  6)   10( 4  4)    7( 5  2)    6( 6  0)    3( 1  7)    6( 2  5)    9( 3  3)   | \\
\verb|  6( 4  1)    5( 0  6)    6( 1  4)    8( 2  2)    2( 3  0)    2( 0  3)    2( 1  1)    2( 0  0)   | \\
$\l\{2^5\r\}\;;\;\;d=19404$ \\
\verb|  1(10  4)    1(12  0)    1( 7  7)    1( 8  5)    2( 9  3)    2(10  1)    1( 4 10)    1( 5  8)   | \\
\verb|  4( 6  6)    4( 7  4)    6( 8  2)    2( 3  9)    4( 4  7)    8( 5  5)    7( 6  3)    6( 7  1)   | \\
\verb|  1( 0 12)    2( 1 10)    6( 2  8)    7( 3  6)   14( 4  4)    9( 5  2)    6( 6  0)    6( 1  7)   | \\
\verb|  9( 2  5)   11( 3  3)    7( 4  1)    6( 0  6)    7( 1  4)   11( 2  2)    1( 3  0)    1( 0  3)   | \\
\verb|  4( 1  1)    1( 0  0) | \\
$\l\{2^6\r\}\;;\;\;d=13860$ \\
\verb|  1(12  0)    1( 8  5)    1( 9  3)    1( 4 10)    1( 5  8)    3( 6  6)    2( 7  4)    4( 8  2)   | \\
\verb|  2( 3  9)    3( 4  7)    5( 5  5)    6( 6  3)    3( 7  1)    1( 0 12)    1( 1 10)    5( 2  8)   | \\
\verb|  7( 3  6)   10( 4  4)    6( 5  2)    5( 6  0)    1( 0  9)    5( 1  7)    7( 2  5)    9( 3  3)   | \\
\verb|  6( 4  1)    6( 0  6)    6( 1  4)    8( 2  2)    2( 3  0)    2( 0  3)    2( 1  1)    2( 0  0)   | \\
$\l\{2^7\r\}\;;\;\;d=4950$ \\
\verb|  1( 6  6)    1( 8  2)    1( 3  9)    1( 4  7)    2( 5  5)    2( 6  3)    1( 7  1)    1( 0 12)   | \\
\verb|  3( 2  8)    3( 3  6)    5( 4  4)    2( 5  2)    3( 6  0)    2( 1  7)    3( 2  5)    5( 3  3)   | \\
\verb|  2( 4  1)    4( 0  6)    3( 1  4)    5( 2  2)    1( 3  0)    1( 0  3)    1( 1  1)    2( 0  0)   | \\
$\l\{2^8\r\}\;;\;\;d=825$ \\
\verb|  1( 2  8)    2( 4  4)    1( 6  0)    1( 1  7)    1( 2  5)    1( 3  3)    1( 4  1)    1( 0  6)   | \\
\verb|  1( 1  4)    2( 2  2)    1( 1  1) | \\
$\l\{2^9\r\}\;;\;\;d=55$ \\
\verb|  1( 0  6)    1( 2  2) | \\
$\l\{2^{10}\r\}\;;\;\;d=1$ \\
\verb|   1( 0  0) | \\
\hline
\end{tabular}
}}
\end{table}
\begin{table}
\caption{Results for $U(15) \supset SU(3)$ reductions with $\eta=4$ for identical fermions with spin $S=0$, i.e. for all $\{2^r\}$ irreps of $U(15)$ with $r=0$ to $15$. The 'd' in the table gives dimension of the irrep $\l\{2^r\r\}$. In the table, $\Gamma$ in $\Gamma(\la\;\;\mu)$ is the multiplicity of the $SU(3)$ irrep $(\la\;\;\mu)$. The irreps with $\Gamma=0$ are not shown as they will not exist in the reductions. Listed only are those $(\la , \mu)$ irreps with $2\la+\mu=\epsilon_H$, $\epsilon_H-3$ and $\epsilon_H-6$.}
{\tiny{
\begin{tabular}{l}
\hline
$\l\{0\r\}\;;\;\;d=1$ \\
\verb|   1( 0  0) | \\
$\l\{2\r\}\;;\;\;d=120$ \\
\verb|   1( 8  0)     1( 4  2) | \\
$\l\{2^2\r\}\;;\;\;d=4200$ \\
\verb|   1(12  2)    1(11  1)    2( 8  4)    1( 9  2)    1(10  0) | \\
$\l\{2^3\r\}\;;\;\;d=63700$ \\
\verb|   1(18  0)    1(15  3)    1(12  6)    1(13  4)    3(14  2) | \\
$\l\{2^4\r\}\;;\;\;d=496860$ \\
\verb|    1(18  4)    1(20  0)    1(16  5)    2(17  3)    1(18  1)    1(13  8)    4(14  6)    4(15  4)   | \\
\verb|  6(16  2)    1(17  0) | \\
$\l\{2^5\r\}\;;\;\;d=2186184$ \\
\verb| 1(20  4)    1(22  0)    1(17  7)    1(18  5)    2(19  3)    2(20  1)    1(14 10)    2(15  8)   | \\
\verb|  6(16  6)    6(17  4)    8(18  2)    1(19  0) | \\
$\l\{2^6\r\}\;;\;\;d=5725720$ \\
\verb| 1(24  0)    1(20  5)    1(21  3)    1(16 10)    1(17  8)    4(18  6)    3(19  4)    5(20  2)  | \\
$\l\{2^7\r\}\;;\;\;d=9202050$ \\
\verb|   1(20  6)    1(22  2)    1(17  9)    2(18  7)    3(19  5)    3(20  3)    2(21  1)    2(14 12)   | \\
\verb|  3(15 10)    9(16  8)   10(17  6)   14(18  4)    8(19  2)    6(20  0) | \\
$\l\{2^8\r\}\;;\;\;d=9202050$ \\
\verb|   1(18  8)    2(20  4)    1(22  0)    1(15 11)    2(16  9)    4(17  7)    5(18  5)    5(19  3)   | \\
\verb|  3(20  1)    2(12 14)    3(13 12)    9(14 10)   12(15  8)   20(16  6)   16(17  4)   18(18  2)   | \\
\verb|  3(19  0)  | \\
$\l\{2^9\r\}\;;\;\;d=5725720$ \\
\verb|   1(18  6)    1(20  2)    1(14 11)    2(15  9)    2(16  7)    4(17  5)    4(18  3)    2(19  1)   | \\
\verb|  1(10 16)    1(11 14)    5(12 12)    6(13 10)   13(14  8)   13(15  6)   17(16  4)    9(17  2)   | \\
\verb|  7(18  0)| \\
$\l\{2^{10}\r\}\;;\;\;d=2186184$ \\
\verb|   1(20  0)    1(15  7)    1(16  5)    1(17  3)    1(10 14)    1(11 12)    4(12 10)    3(13  8)   | \\
\verb|  7(14  6)    4(15  4)    6(16  2)| \\
$\l\{2^{11}\r\}\;;\;\;d=496860$ \\
\verb|    1(12  8)    1(14  4)    1(16  0)    1( 8 13)    2( 9 11)    2(10  9)    4(11  7)    4(12  5)   | \\
\verb|  3(13  3)    2(14  1)    1( 4 18)    1( 5 16)    4( 6 14)    5( 7 12)   11( 8 10)   11( 9  8)   | \\
\verb| 16(10  6)   11(11  4)   12(12  2)    1(13  0) | \\
$\l\{2^{12}\r\}\;;\;\;d=63700$ \\
\verb|   1( 6 12)    2( 8  8)    1( 9  6)    2(10  4)    2(12  0)    1( 3 15)    1( 4 13)    3( 5 11)   | \\
\verb|  4( 6  9)    5( 7  7)    5( 8  5)    5( 9  3)    2(10  1)    1( 0 18)    3( 2 14)    4( 3 12)   | \\
\verb|  8( 4 10)    8( 5  8)   14( 6  6)    8( 7  4)   10( 8  2)    1( 9  0)  | \\
$\l\{2^{13}\r\}\;;\;\;d=4200$ \\
\verb|   1( 2 12)    2( 4  8)    1( 5  6)    2( 6  4)    2( 8  0)    1( 1 11)    1( 2  9)    2( 3  7)   | \\
\verb|  2( 4  5)    2( 5  3)    1( 6  1)    1( 0 10)    1( 1  8)    3( 2  6)    2( 3  4)    3( 4  2) | \\
$\l\{2^{14}\r\}\;;\;\;d=120$ \\
\verb|   1( 0  8)     1( 2  4)     1( 4  0) | \\
$\l\{2^{15}\r\}\;;\;\;d=1$ \\
\verb|   1( 0  0) | \\
\hline
\end{tabular}
}}
\end{table}

\subsection{Results for two column irreps of $U(\can)$; $\can=10,15$}

Using the method described in the previous subsection, we have obtained the reductions for all two column irreps $\l\{2^r\r\}$ for all values of $r$ for $\eta=3$ (then $\can=10$ and $r=0-10$)
and $\eta=4$ (then $\can=15$ and $r=0-15$). Table I gives the
results for $\eta=3$ shell. Note that an irrep $(\la , \mu)$ is said to be of higher weight than the irrep $(\la^\pr , \mu^\pr)$ [then we say $(\la , \mu) > (\la^\pr , \mu^\pr)$] if either of the following is satisfied,
\be
\barr{l}
(i)\;\;(\la , \mu) > (\la^\pr , \mu^\pr)\;\;\mbox{if}\;\;2\la+\mu > 2\la^\pr +\mu^\pr\;,\\
\mbox{or} \\
(ii)\;\;(\la , \mu) > (\la^\pr , \mu^\pr)\;\;\mbox{if}\;\;2\la+\mu = 2\la^\pr +\mu^\pr\;\;\mbox{and}\;\;\mu > \mu^\pr \;.
\earr \label{su4-10a}
\ee 
In all the tables in this paper the $(\la , \mu)$ are arranged in
decreasing weight. Thus, the first $SU(3)$ irrep appearing in
Table 1 for any given $\l\{2^r\r\}$ of $U(10)$ is the hw $SU(3)$ irrep and it is denoted by $(\la_H , \mu_H)$. In the following we will also use $\epsilon_H = 2\la_H +\mu_H$. 
A simple formula for hw irrep, for any $\eta$ and any $\{f\}$ of $U(\can)$ [$\can = (\eta+1)(\eta+2)/2$], is as follows \cite{Ko-book,Ko-hw} 
\be 
\barr{l} \l\{f\r\}_{U((\eta
+1)(\eta +2)/2)} \rightarrow (\lambda_H , \mu_H)_{SU(3)}  \;\;\mbox{with} \\
\lambda_H = \dis\sum_{r=0}^\eta \dis\sum_{x=0}^r (\eta -2r+x) \times 
f_{1+x+\frac{r(r+1)}{2}}\;,\;\;\;\;\mu_H=\dis\sum_{r=0}^\eta \dis\sum_{x=0}^r 
(r-2x) \times f_{1+x+\frac{r(r+1)}{2}}\;. 
\earr 
\label{su4-11} 
\ee 
Results from Eq. (\ref{su4-11}) agree with the hw irrep obtained
from Table I for all $\l\{2^r\r\}$ irreps. Let us mention that,
by extending Eq. (\ref{su4-3}) it is easy to obtain the reductions for $\l\{2^r\r\}$ for $r > \can/2$ from the reductions for $\l\{2^r\r\}$ irreps with $r < \can/2$,
\be
\barr{l}
\l\{2^r\r\} \rightarrow \dis\sum_i x_i (\la_i , \mu_i) \oplus \\
\Rightarrow \l\{2^{\can -r}\r\} \rightarrow \dis\sum_i x_i (\mu_i , \la_i) \oplus\;.
\earr \label{su4-12}
\ee
Now, more importantly the nhw irrep as given in Table 1 of \cite{Bona-5} (also Table 2
\cite{Bona-7}) agree with those given here in Table I here if we ignore the irreps $(\la , \mu)$ with $\la$ odd or $\mu$ odd. Also, in \cite{Bona-5,Bona-7} tables, the multiplicity of nhw is not shown . Note that for hw irrep the multiplicity is always one but for nhw the multiplicity is in many situations more than one as seen from Table I. Let us add that the results in Table I are verified using the dimension
formulas in Eq. (\ref{su4-10}). In the table listed are also the dimension of a given $\l\{2^r\r\}$ irrep. Going further, in Table II listed are the reductions for two column irreps for $\eta=4$. As the current interest is only in hw and nhw irreps, in Table II listed only are those $(\la , \mu)$ irreps with $2\la+\mu=\epsilon_H$, $\epsilon_H-3$ and $\epsilon_H-6$ (complete reductions are available with the author). Once again it is seen from Table II that the hw irrep and the nhw irrep for all $\l\{2^r\r\}$ irreps agree with those listed in \cite{Bona-5,Bona-7} provided
we ignore the irreps $(\la , \mu)$ with $\la$ odd or $\mu$ odd. This is the case with some of the $\l\{2^r\r\}$ irreps as seen from Table II. Also, note that the nhw irrep in many cases has multiplicity more than one. It is important to note that the hw irrep always has both $\la$ and $\mu$ even and therefore generates a $K=0$ band with all $L$ even (then, all $J$ even as we are considering only $S=0$ situation). This follows from the general result \cite{Ell-58a,Ell-58b},
\be
\barr{rcl}
(\lambda \mu) \longrightarrow L: & & \\
K & = & min(\lambda,\mu),\; min(\lambda,\mu) -2,\; \cdots , 0\;\mbox{or}\;1, \\
L &=& K,\; K+1, \;K+2,\; \cdots , \; K + max(\lambda,\mu)\;\; for\;\;
K \neq 0\;, \\
L & = & max(\lambda,\mu),\;max(\lambda,\mu)-2,\; \cdots , 0\;\mbox{or}\; 
1\;\;for\;\; K = 0\;.
\earr \label{su3-l}
\ee
As seen from Eq. (\ref{su3-l}), irreps with $\la$ or $\mu$ odd will not generate a $K=0$ band with $L$ even. This appears to be the reason for ignoring them in \cite{Bona-5,Bona-7}. However, irreps with $\la$ or $\mu$ odd do generate $2^+$, $4^+$ and other even $L^\pi$ states and these can mix with the corresponding $L^\pi$ states from the hw $SU(3)$ irrep.

\begin{table}
\caption{{\scriptsize{Low-lying $SU(3)$ irreps (giving hw irrep, nhw irrep, nnhw irrep etc.) $(\la , \mu)$ for $^{64,68,70,72,74,76}$Ge isotopes. In the table results are shown as $\Gamma(\la\;\;\mu)$ where $\Gamma$ is the multiplicity of the $SU(3)$ irrep $(\la\;\;\mu)$. The irreps with $\Gamma=0$ are not shown as they will not exist. Listed only are those $(\la , \mu)$ irreps with $2\la+\mu=\epsilon_H$, $\epsilon_H-3$ and $\epsilon_H-6$. Shown in the table are also number of valence nucleons $m$, isospin $T$ and the $U(10)$ irrep $\l\{f\r\}$. See text for further details.}}}
{\scriptsize{
\begin{tabular}{l}
\hline
$^{64}$Ge $\;:\;\;(m=8,T=0)\;:\;\;\l\{f\r\}=\l\{4^2\r\}$ \\
\verb|  1(16  4)   1(15  3)   1(16  1)   2(12  6)   2(13  4)   3(14  2)   1(15  0) | \\
$^{66}$Ge $\;:\;\;(m=10,T=1)\;:\;\;\l\{f\r\}=\l\{4^2,2\r\}$ \\
\verb|  1(20  2)   1(17  5)   1(18  3)   2(19  1)   1(14  8)   2(15  6)   6(16  4) | \\
\verb|  5(17  2)   4(18  0) | \\
$^{68}$Ge $\;:\;\;(m=12,T=2)\;:\;\;\l\{f\r\}=\l\{4^2,2^2\r\}$ \\
\verb|  1(18  6)   1(19  4)   2(20  2)   2(16  7)   5(17  5)   6(18  3)   5(19  1) | \\ 
\verb|  1(13 10)   7(14  8)  13(15  6)  23(16  4)  18(17  2)  10(18  0) | \\
$^{70}$Ge $\;:\;\;(m=14,T=3)\;:\;\;\l\{f\r\}=\l\{4^2,2^3\r\}$ \\
\verb|  1(18  6)   1(19  4)   2(20  2)   1(15  9)   3(16  7)   7(17  5)   8(18  3) | \\
\verb|  6(19  1)   1(12 12)   3(13 10)  12(14  8)  20(15  6)  32(16  4)  25(17  2) | \\
\verb| 14(18  0) | \\
$^{72}$Ge $\;:\;\;(m=16,T=4)\;:\;\;\l\{f\r\}=\l\{4^2,2^4\r\}$ \\
\verb|  1(20  2)   1(16  7)   2(17  5)   3(18  3)   3(19  1)   1(12 12)   2(13 10) | \\
\verb|  6(14  8)  10(15  6)  17(16  4)  12(17  2)   8(18  0) | \\
$^{74}$Ge $\;:\;\;(m=18,T=5)\;:\;\;\l\{f\r\}=\l\{4^2,2^5\r\}$ \\
\verb|  1(14  8)   1(15  6)   2(16  4)   1(17  2)   1(18  0)   1(11 11)   3(12  9) | \\
\verb|  7(13  7)   9(14  5)   9(15  3)   6(16  1)   1( 8 14)   2( 9 12)  11(10 10) | \\
\verb| 19(11  8)  33(12  6)  34(13  4)  31(14  2)   8(15  0) | \\
$^{76}$Ge $\;:\;\;(m=20,T=6)\;:\;\;\l\{f\r\}=\l\{4^2,2^6\r\}$ \\
\verb|  1(10 10)   1(11  8)   3(12  6)   2(13  4)   3(14  2)   1( 8 11)   4( 9  9) | \\
\verb|  7(10  7)  10(11  5)  10(12  3)   6(13  1)   3( 6 12)   7( 7 10)  19( 8  8) | \\
\verb| 25( 9  6)  34(10  4)  23(11  2)  13(12  0) | \\
\hline
\end{tabular}
}}
\end{table}
\begin{table}
\caption{{\scriptsize{Low-lying $SU(3)$ irreps (giving hw irrep, nhw irrep, nnhw irrep etc.) $(\la , \mu)$ for $^{70,72,74,76,78,80,82}$Se isotopes. In the table results are shown as $\Gamma(\la\;\;\mu)$ where $\Gamma$ is the multiplicity of the $SU(3)$ irrep $(\la\;\;\mu)$. The irreps with $\Gamma=0$ are not shown as they will not exist. Listed only are those $(\la , \mu)$ irreps with $2\la+\mu=\epsilon_H$, $\epsilon_H-3$ and $\epsilon_H-6$. Shown in the table are also number of valence nucleons $m$, isospin $T$ and the $U(10)$ irrep $\l\{f\r\}$. In addition $\l\{f_h\r\}$ is also
given when appropriate. See text for further details.}}}
{\scriptsize{
\begin{tabular}{l}
\hline
$^{70}$Se $\;:\;\;(m=14,T=1)\;:\;\;\l\{f\r\}=\l\{4^3,2\r\}$ \\
\verb|  1(22  4)   1(24  0)   1(19  7)   2(20  5)   3(21  3)   2(22  1)   1(16 10) | \\
\verb|  3(17  8)   9(18  6)   9(19  4)  12(20  2)   2(21  0) | \\
$^{72}$Se $\;:\;\;(m=16,T=2)\;:\;\;\l\{f\r\}=\l\{4^3,2^2\r\}$ \\ \verb|  1(22  4)   1(24  0)   2(19  7)   3(20  5)   4(21  3)   3(22  1)   3(16 10) | \\
\verb|  6(17  8)  15(18  6)  16(19  4)  19(20  2)   3(21  0) | \\
$^{74}$Se $\;:\;\;(m=18,T=3)\;:\;\;\l\{f\r\}=\l\{4^3,2^3\r\}$ \\  
\verb|  1(24  0)   1(20  5)   2(21  3)   1(22  1)   1(16 10)   2(17  8)   7(18  6) | \\
\verb|  7(19  4)  11(20  2)   1(21  0) | \\
$^{76}$Se $\;:\;\;(m=20,T=4)\;:\;\;\l\{f\r\}=\l\{4^3,2^4\r\}$ \\ 
\verb|  1(18  6)   1(20  2)   2(15  9)   3(16  7)   5(17  5)   5(18  3)   3(19  1) | \\
\verb|  3(12 12)   6(13 10)  17(14  8)  21(15  6)  29(16  4)  18(17  2)  11(18  0) | \\
$^{78}$Se $\;:\;\;(m=22,T=5)\;:\;\;\l\{f\r\}=\l\{4^3,2^5\r\},\;
\l\{f_h\r\}=\l\{4^2,2^5\r\}$ \\ 
\verb|  1(14  8)   2(16  4)   1(18  0)   1(11 11)   2(12  9)   5(13  7)   6(14  5) | \\ 
\verb|  6(15  3)   4(16  1)   1( 8 14)   3( 9 12)  11(10 10)  16(11  8)  29(12  6) | \\
\verb| 26(13  4)  27(14  2)   5(15  0) | \\
$^{80}$Se $\;:\;\;(m=24,T=6)\;:\;\;\l\{f\r\}=\l\{4^3,2^6\r\},\;
\l\{f_h\r\}=\l\{4,2^6\r\}$ \\  
\verb|  1(12  6)   1(14  2)   1( 9  9)   1(10  7)   3(11  5)   3(12  3)   2(13  1) | \\
\verb|  1( 6 12)   2( 7 10)   7( 8  8)   9( 9  6)  14(10  4)   9(11  2)   6(12  0) | \\
$^{82}$Se $\;:\;\;(m=26,T=7)\;:\;\;\l\{f\r\}=\l\{4^3,2^7\r\},\;
\l\{f_h\r\}=\l\{2^7\r\}$ \\  
\verb|  1(12  0)   1( 9  3)   1( 6  6)   1( 7  4)   3( 8  2) | \\
\hline
\end{tabular}
}}
\end{table}
\begin{table}
\caption{{\scriptsize{Low-lying $SU(3)$ irreps (giving hw irrep, nhw irrep, nnhw irrep etc.) $(\la , \mu)$ for $^{72,74,76,78,80,82}$Kr isotopes. In the table results are shown as $\Gamma(\la\;\;\mu)$ where $\Gamma$ is the multiplicity of the $SU(3)$ irrep $(\la\;\;\mu)$. The irreps with $\Gamma=0$ are not shown as they will not exist. Listed only are those $(\la , \mu)$ irreps with $2\la+\mu=\epsilon_H$, $\epsilon_H-3$ and $\epsilon_H-6$. Shown in the table are also number of valence nucleons $m$, isospin $T$ and the $U(10)$ irrep $\l\{f\r\}$. In addition $\l\{f_h\r\}$ is also
given when appropriate. See text for further details.}}}
{\scriptsize{
\begin{tabular}{l}
\hline
$^{72}$Kr $\;:\;\;(m=16,T=0)\;:\;\;\l\{f\r\}=\l\{4^4\r\}$ \\
\verb|  1(20  8)   1(22  4)   1(24  0)   1(18  9)   2(19  7)   3(20  5)   3(21  3) | \\
\verb|  1(22  1)   1(15 12)   3(16 10)   6(17  8)  11(18  6)   9(19  4)  10(20  2) | \\
\verb|  2(21  0) | \\
$^{74}$Kr $\;:\;\;(m=18,T=1)\;:\;\;\l\{f\r\}=\l\{4^4,2\r\}$ \\
\verb|  1(20  8)   1(21  6)   2(22  4)   1(24  0)   1(17 11)   3(18  9)   6(19  7) | \\
\verb|  8(20  5)   8(21  3)   5(22  1)   1(14 14)   3(15 12)  11(16 10)  19(17  8) | \\
\verb| 32(18  6)  32(19  4)  30(20  2)   7(21  0) | \\
$^{76}$Kr $\;:\;\;(m=20,T=2)\;:\;\;\l\{f\r\}=\l\{4^4,2^2\r\}$ \\ 
\verb|  1(22  4)   1(24  0)   1(18  9)   2(19  7)   4(20  5)   5(21  3)   3(22  1) | \\
\verb|  1(14 14)   2(15 12)   7(16 10)  12(17  8)  22(18  6)  22(19  4)  24(20  2) | \\ 
\verb|  5(21  0) | \\
$^{78}$Kr $\;:\;\;(m=22,T=3)\;:\;\;\l\{f\r\}=\l\{4^4,2^3\r\},\;
\l\{f_h\r\}=\l\{4^3,2^3\r\}$ \\  
\verb|  1(16 10)   1(17  8)   3(18  6)   2(19  4)   3(20  2)   1(13 13)   4(14 11) | \\
\verb|  9(15  9)  14(16  7)  18(17  5)  17(18  3)  10(19  1)   1(10 16)   4(11 14) | \\
\verb| 16(12 12)  30(13 10)  56(14  8)  69(15  6)  78(16  4)  54(17  2)  26(18  0) | \\
$^{80}$Kr $\;:\;\;(m=24,T=4)\;:\;\;\l\{f\r\}=\l\{4^4,2^4\r\},\;
\l\{f_h\r\}=\l\{4^2,2^4\r\}$ \\
\verb|  1(12 12)   1(13 10)   4(14  8)   3(15  6)   6(16  4)   2(17  2)   3(18  0) | \\ 
\verb|  2(10 13)   6(11 11)  12(12  9)  19(13  7)  23(14  5)  21(15  3)  13(16  1) | \\
\verb|  1( 7 16)   6( 8 14)  16( 9 12)  38(10 10)  57(11  8)  82(12  6)  78(13  4) | \\
\verb| 67(14  2)  19(15  0) | \\
$^{82}$Kr $\;:\;\;(m=26,T=5)\;:\;\;\l\{f\r\}=\l\{4^4,2^5\r\},\;
\l\{f_h\r\}=\l\{4,2^5\r\}$ \\
\verb|  1(10 10)   1(11  8)   3(12  6)   2(13  4)   3(14  2)   1( 8 11)   4( 9  9) | \\
\verb|  7(10  7)  10(11  5)  10(12  3)   6(13  1)   1( 5 14)   4( 6 12)   9( 7 10) | \\
\verb| 21( 8  8)  28( 9  6)  35(10  4)  25(11  2)  12(12  0) | \\
\hline
\end{tabular}
}}
\end{table}
\begin{table}
\caption{{\scriptsize{Low-lying $SU(3)$ irreps (giving hw irrep, nhw irrep, nnhw irrep etc.) $(\la , \mu)$ for $^{74,76,78,80,82,84}$Sr isotopes. In the table results are shown as $\Gamma(\la\;\;\mu)$ where $\Gamma$ is the multiplicity of the $SU(3)$ irrep $(\la\;\;\mu)$. The irreps with $\Gamma=0$ are not shown as they will not exist. Listed only are those $(\la , \mu)$ irreps with $2\la+\mu=\epsilon_H$, $\epsilon_H-3$ and $\epsilon_H-6$. Shown in the table are also number of valence nucleons $m$, isospin $T$ and the $U(10)$ irrep $\l\{f\r\}$. In addition $\l\{f_h\r\}$ is also
given when appropriate. See text for further details.}}}
{\scriptsize{
\begin{tabular}{l}
\hline
$^{74}$Sr $\;:\;\;(m=18,T=1)\;:\;\;\l\{f\r\}=\l\{4^4,2\r\}$ \\
\verb|  1(20  8)   1(21  6)   2(22  4)   1(24  0)   1(17 11)   3(18  9)   6(19  7) | \\
\verb|  8(20  5)   8(21  3)   5(22  1)   1(14 14)   3(15 12)  11(16 10)  19(17  8) | \\
\verb| 32(18  6)  32(19  4)  30(20  2)   7(21  0) | \\
$^{76}$Sr $\;:\;\;(m=20,T=0)\;:\;\;\l\{f\r\}=\l\{4^5\r\}$ \\ 
\verb|  1(20  8)   1(22  4)   1(24  0)   1(17 11)   1(18  9)   3(19  7)   4(20  5) | \\
\verb|  3(21  3)   2(22  1)   1(14 14)   2(15 12)   6(16 10)   8(17  8)  14(18  6) | \\ 
\verb| 12(19  4)  14(20  2)   2(21  0) | \\
$^{78}$Sr $\;:\;\;(m=22,T=1)\;:\;\;\l\{f\r\}=\l\{4^5,2\r\},\;
\l\{f_h\r\}=\l\{4^4,2\r\}$ \\ 
\verb|  1(22  4)   1(24  0)   1(18  9)   2(19  7)   3(20  5)   4(21  3)   3(22  1) | \\
\verb|  1(14 14)   2(15 12)   6(16 10)   9(17  8)  17(18  6)  16(19  4)  18(20  2) | \\
\verb|  3(21  0) | \\
$^{80}$Sr $\;:\;\;(m=24,T=2)\;:\;\;\l\{f\r\}=\l\{4^5,2^2\r\},\;
\l\{f_h\r\}=\l\{4^3,2^2\r\}$ \\  
\verb|  1(16 10)   1(17  8)   3(18  6)   2(19  4)   3(20  2)   2(13 13)   4(14 11) | \\
\verb|  9(15  9)  14(16  7)  17(17  5)  15(18  3)  10(19  1)   3(10 16)   6(11 14) | \\
\verb| 19(12 12)  32(13 10)  55(14  8)  63(15  6)  72(16  4)  47(17  2)  23(18  0) | \\
$^{82}$Sr $\;:\;\;(m=26,T=3)\;:\;\;\l\{f\r\}=\l\{4^5,2^3\r\},\;
\l\{f_h\r\}=\l\{4^2,2^3\r\}$ \\ 
\verb|  1(12 12)   1(13 10)   4(14  8)   3(15  6)   6(16  4)   2(17  2)   3(18  0) | \\
\verb|  1( 9 15)   3(10 13)   8(11 11)  14(12  9)  21(13  7)  24(14  5)  22(15  3) | \\
\verb| 13(16  1)   1( 6 18)   3( 7 16)  12( 8 14)  23( 9 12)  48(10 10)  65(11  8) | \\
\verb| 89(12  6)  80(13  4)  69(14  2)  18(15  0) | \\
$^{84}$Sr $\;:\;\;(m=28,T=4)\;:\;\;\l\{f\r\}=\l\{4^5,2^4\r\},\;
\l\{f_h\r\}=\l\{4,2^4\r\}$ \\ 
\verb|  1(10 10)   1(11  8)   3(12  6)   2(13  4)   3(14  2)   1( 7 13)   2( 8 11) | \\
\verb|  6( 9  9)   9(10  7)  12(11  5)  11(12  3)   7(13  1)   1( 4 16)   2( 5 14) | \\
\verb|  8( 6 12)  14( 7 10)  28( 8  8)  33( 9  6)  40(10  4)  27(11  2)  14(12  0) | \\
\hline
\end{tabular}
}}
\end{table}
\begin{table}
\caption{{\scriptsize{Low-lying $SU(3)$ irreps (giving hw irrep, nhw irrep, nnhw irrep etc.) $(\la , \mu)$ for $^{76,78,80,82,84,86}$Zr isotopes. In the table results are shown as $\Gamma(\la\;\;\mu)$ where $\Gamma$ is the multiplicity of the $SU(3)$ irrep $(\la\;\;\mu)$. The irreps with $\Gamma=0$ are not shown as they will not exist. Listed only are those $(\la , \mu)$ irreps with $2\la+\mu=\epsilon_H$, $\epsilon_H-3$ and $\epsilon_H-6$. Shown in the table are also number of valence nucleons $m$, isospin $T$ and the $U(10)$ irrep $\l\{f\r\}$. In addition $\l\{f_h\r\}$ is also
given when appropriate. See text for further details.}}}
{\scriptsize{
\begin{tabular}{l}
\hline
$^{76}$Zr $\;:\;\;(m=20,T=2)\;:\;\;\l\{f\r\}=\l\{4^4,2^2\r\}$ \\
\verb|  1(22  4)   1(24  0)   1(18  9)   2(19  7)   4(20  5)   5(21  3)   3(22  1) | \\
\verb|  1(14 14)   2(15 12)   7(16 10)  12(17  8)  22(18  6)  22(19  4)  24(20  2) | \\
\verb|  5(21  0) | \\
$^{78}$Zr $\;:\;\;(m=22,T=1)\;:\;\;\l\{f\r\}=\l\{4^5,2\r\},\;
\l\{f_h\r\}=\l\{4^4,2\r\}$ \\  
\verb|  1(22  4)   1(24  0)   1(18  9)   2(19  7)   3(20  5)   4(21  3)   3(22  1) | \\
\verb|  1(14 14)   2(15 12)   6(16 10)   9(17  8)  17(18  6)  16(19  4)  18(20  2) | \\
\verb|  3(21  0) | \\
$^{80}$Zr $\;:\;\;(m=24,T=0)\;:\;\;\l\{f\r\}=\l\{4^6\r\},\;
\l\{f_h\r\}=\l\{4^4\r\}$ \\  
\verb|  1(24  0)   1(20  5)   1(21  3)   1(16 10)   1(17  8)   3(18  6)   2(19  4) | \\
\verb|  4(20  2) | \\
$^{82}$Zr $\;:\;\;(m=26,T=1)\;:\;\;\l\{f\r\}=\l\{4^6,2\r\},\;
\l\{f_h\r\}=\l\{4^3,2\r\}$ \\  
\verb|  1(18  6)   1(20  2)   1(14 11)   2(15  9)   3(16  7)   4(17  5)   4(18  3) | \\
\verb|  2(19  1)   1(10 16)   2(11 14)   6(12 12)   9(13 10)  16(14  8)  17(15  6) | \\
\verb| 20(16  4)  12(17  2)   7(18  0) | \\
$^{84}$Zr $\;:\;\;(m=28,T=2)\;:\;\;\l\{f\r\}=\l\{4^6,2^2\r\},\;
\l\{f_h\r\}=\l\{4^2,2^2\r\}$ \\   
\verb|  1(14  8)   2(16  4)   1(18  0)   1(10 13)   2(11 11)   4(12  9)   6(13  7) | \\
\verb|  7(14  5)   6(15  3)   4(16  1)   1( 6 18)   2( 7 16)   7( 8 14)  11( 9 12) | \\
\verb| 22(10 10)  26(11  8)  36(12  6)  29(13  4)  27(14  2)   5(15  0) | \\
$^{86}$Zr $\;:\;\;(m=30,T=3)\;:\;\;\l\{f\r\}=\l\{4^6,2^3\r\},\;
\l\{f_h\r\}=\l\{4,2^3\r\}$ \\   
\verb|  1(12  6)   1(14  2)   1( 8 11)   2( 9  9)   3(10  7)   4(11  5)   4(12  3) | \\
\verb|  2(13  1)   1( 4 16)   2( 5 14)   6( 6 12)   9( 7 10)  16( 8  8)  17( 9  6) | \\
\verb| 20(10  4)  12(11  2)   7(12  0) | \\
\hline
\end{tabular}
}}
\end{table}
\begin{table}
\caption{{\scriptsize{Low-lying $SU(3)$ irreps (giving hw irrep, nhw irrep, nnhw irrep etc.) $(\la , \mu)$ for $^{78,80,82,84,86,88}$Mo isotopes. In the table results are shown as $\Gamma(\la\;\;\mu)$ where $\Gamma$ is the multiplicity of the $SU(3)$ irrep $(\la\;\;\mu)$. The irreps with $\Gamma=0$ are not shown as they will not exist. Listed only are those $(\la , \mu)$ irreps with $2\la+\mu=\epsilon_H$, $\epsilon_H-3$ and $\epsilon_H-6$. Shown in the table are also number of valence nucleons $m$, isospin $T$ and the $U(10)$ irrep $\l\{f\r\}$. In addition $\l\{f_h\r\}$ is also
given when appropriate. See text for further details.}}}
{\scriptsize{
\begin{tabular}{l}
\hline
$^{78}$Mo $\;:\;\;(m=22,T=3)\;:\;\;\l\{f\r\}=\l\{4^4,2^3\r\},\;
\l\{f_h\r\}=\l\{4^3,2^3\r\}$ \\   
\verb|  1(16 10)   1(17  8)   3(18  6)   2(19  4)   3(20  2)   1(13 13)   4(14 11) | \\
\verb|  9(15  9)  14(16  7)  18(17  5)  17(18  3)  10(19  1)   1(10 16)   4(11 14) | \\
\verb| 16(12 12)  30(13 10)  56(14  8)  69(15  6)  78(16  4)  54(17  2)  26(18  0) | \\
$^{80}$Mo $\;:\;\;(m=24,T=2)\;:\;\;\l\{f\r\}=\l\{4^5,2^2\r\},\;
\l\{f_h\r\}=\l\{4^3,2^2\r\}$ \\ 
\verb|  1(16 10)   1(17  8)   3(18  6)   2(19  4)   3(20  2)   2(13 13)   4(14 11) | \\
\verb|  9(15  9)  14(16  7)  17(17  5)  15(18  3)  10(19  1)   3(10 16)   6(11 14) | \\
\verb| 19(12 12)  32(13 10)  55(14  8)  63(15  6)  72(16  4)  47(17  2)  23(18  0) | \\
$^{82}$Mo $\;:\;\;(m=26,T=1)\;:\;\;\l\{f\r\}=\l\{4^6,2\r\},\;
\l\{f_h\r\}=\l\{4^3,2\r\}$ \\  
\verb|  1(18  6)   1(20  2)   1(14 11)   2(15  9)   3(16  7)   4(17  5)   4(18  3) | \\
\verb|  2(19  1)   1(10 16)   2(11 14)   6(12 12)   9(13 10)  16(14  8)  17(15  6) | \\
\verb| 20(16  4)  12(17  2)   7(18  0) | \\
$^{84}$Mo $\;:\;\;(m=28,T=0)\;:\;\;\l\{f\r\}=\l\{4^7\r\},\;
\l\{f_h\r\}=\l\{4^3\r\}$ \\  
\verb|  1(12 12)   1(14  8)   1(15  6)   1(16  4)   1(18  0)   1( 9 15)   1(10 13) | \\
\verb|  3(11 11)   4(12  9)   4(13  7)   4(14  5)   4(15  3)   2(16  1)   1( 6 18) | \\
\verb|  1( 7 16)   5( 8 14)   7( 9 12)  12(10 10)  13(11  8)  17(12  6)  12(13  4) | \\
\verb| 11(14  2)   2(15  0) | \\
$^{86}$Mo $\;:\;\;(m=30,T=1)\;:\;\;\l\{f\r\}=\l\{4^7,2\r\},\;
\l\{f_h\r\}=\l\{4^2,2\r\}$ \\   
\verb|  1( 8 14)   1( 9 12)   3(10 10)   2(11  8)   4(12  6)   2(13  4)   3(14  2) | \\
\verb|  1( 5 17)   2( 6 15)   5( 7 13)   8( 8 11)  12( 9  9)  14(10  7)  14(11  5) | \\
\verb| 11(12  3)   6(13  1)   1( 2 20)   1( 3 18)   6( 4 16)  10( 5 14)  22( 6 12) | \\
\verb| 29( 7 10)  42( 8  8)  40( 9  6)  42(10  4)  24(11  2)  13(12  0) | \\
$^{88}$Mo $\;:\;\;(m=32,T=2)\;:\;\;\l\{f\r\}=\l\{4^7,2^2\r\},\;
\l\{f_h\r\}=\l\{4,2^2\r\}$ \\   
\verb|  1( 6 12)   1( 7 10)   3( 8  8)   2( 9  6)   3(10  4)   1(11  2)   1(12  0) | \\
\verb|  1( 3 15)   2( 4 13)   5( 5 11)   7( 6  9)  10( 7  7)  10( 8  5)   9( 9  3) | \\
\verb|  5(10  1)   1( 0 18)   1( 1 16)   5( 2 14)   8( 3 12)  17( 4 10)  21( 5  8) | \\
\verb| 29( 6  6)  24( 7  4)  21( 8  2)   5( 9  0) | \\
\hline
\end{tabular}
}}
\end{table}
\begin{table}
\caption{{\scriptsize{Low-lying $SU(3)$ irreps (giving hw irrep, nhw irrep, nnhw irrep etc.) $(\la , \mu)$ for $^{80,82,84,86,88,90}$Ru isotopes. In the table results are shown as $\Gamma(\la\;\;\mu)$ where $\Gamma$ is the multiplicity of the $SU(3)$ irrep $(\la\;\;\mu)$. The irreps with $\Gamma=0$ are not shown as they will not exist. Listed only are those $(\la , \mu)$ irreps with $2\la+\mu=\epsilon_H$, $\epsilon_H-3$ and $\epsilon_H-6$. Shown in the table are also number of valence nucleons $m$, isospin $T$ and the $U(10)$ irrep $\l\{f\r\}$. In addition $\l\{f_h\r\}$ is also
given when appropriate. See text for further details.}}}
{\scriptsize{
\begin{tabular}{l}
\hline
$^{80}$Ru $\;:\;\;(m=24,T=4)\;:\;\;\l\{f\r\}=\l\{4^4,2^4\r\},\;
\l\{f_h\r\}=\l\{4^2,2^4\r\}$ \\   
\verb|  1(12 12)   1(13 10)   4(14  8)   3(15  6)   6(16  4)   2(17  2)   3(18  0) | \\
\verb|  2(10 13)   6(11 11)  12(12  9)  19(13  7)  23(14  5)  21(15  3)  13(16  1) | \\
\verb|  1( 7 16)   6( 8 14)  16( 9 12)  38(10 10)  57(11  8)  82(12  6)  78(13  4) | \\
\verb| 67(14  2)  19(15  0) | \\
$^{82}$Ru $\;:\;\;(m=26,T=3)\;:\;\;\l\{f\r\}=\l\{4^5,2^3\r\},\;
\l\{f_h\r\}=\l\{4^2,2^3\r\}$ \\    
\verb|  1(12 12)   1(13 10)   4(14  8)   3(15  6)   6(16  4)   2(17  2)   3(18  0) | \\
\verb|  1( 9 15)   3(10 13)   8(11 11)  14(12  9)  21(13  7)  24(14  5)  22(15  3) | \\
\verb| 13(16  1)   1( 6 18)   3( 7 16)  12( 8 14)  23( 9 12)  48(10 10)  65(11  8) | \\
\verb| 89(12  6)  80(13  4)  69(14  2)  18(15  0) | \\
$^{84}$Ru $\;:\;\;(m=28,T=2)\;:\;\;\l\{f\r\}=\l\{4^6,2^2\r\},\;
\l\{f_h\r\}=\l\{4^2,2^2\r\}$ \\     
\verb|  1(14  8)   2(16  4)   1(18  0)   1(10 13)   2(11 11)   4(12  9)   6(13  7) | \\
\verb|  7(14  5)   6(15  3)   4(16  1)   1( 6 18)   2( 7 16)   7( 8 14)  11( 9 12) | \\
\verb| 22(10 10)  26(11  8)  36(12  6)  29(13  4)  27(14  2)   5(15  0) | \\
$^{86}$Ru $\;:\;\;(m=30,T=1)\;:\;\;\l\{f\r\}=\l\{4^7,2\r\},\;
\l\{f_h\r\}=\l\{4^2,2\r\}$ \\     
\verb|  1( 8 14)   1( 9 12)   3(10 10)   2(11  8)   4(12  6)   2(13  4)   3(14  2) | \\
\verb|  1( 5 17)   2( 6 15)   5( 7 13)   8( 8 11)  12( 9  9)  14(10  7)  14(11  5) | \\
\verb| 11(12  3)   6(13  1)   1( 2 20)   1( 3 18)   6( 4 16)  10( 5 14)  22( 6 12) | \\
\verb| 29( 7 10)  42( 8  8)  40( 9  6)  42(10  4)  24(11  2)  13(12  0) | \\
$^{88}$Ru $\;:\;\;(m=32,T=0)\;:\;\;\l\{f\r\}=\l\{4^8\r\},\;
\l\{f_h\r\}=\l\{4^2\r\}$ \\     
\verb|  1( 4 16)   2( 6 12)   1( 7 10)   3( 8  8)   1( 9  6)   3(10  4)   2(12  0) | \\ 
\verb|  1( 3 15)   2( 4 13)   3( 5 11)   4( 6  9)   5( 7  7)   5( 8  5)   4( 9  3) | \\
\verb|  2(10  1)   1( 1 16)   3( 2 14)   4( 3 12)   9( 4 10)   9( 5  8)  12( 6  6) | \\
\verb|  9( 7  4)   9( 8  2)   1( 9  0) | \\
$^{90}$Ru $\;:\;\;(m=34,T=1)\;:\;\;\l\{f\r\}=\l\{4^8,2\r\},\;
\l\{f_h\r\}=\l\{4,2\r\}$ \\     
\verb|  1( 2 14)   1( 3 12)   3( 4 10)   2( 5  8)   4( 6  6)   2( 7  4)   3( 8  2) | \\
\verb|  1( 1 13)   2( 2 11)   4( 3  9)   5( 4  7)   6( 5  5)   5( 6  3)   3( 7  1) | \\
\verb|  2( 0 12)   4( 1 10)   8( 2  8)   9( 3  6)  11( 4  4)   7( 5  2)   4( 6  0) | \\
\hline
\end{tabular}
}}
\end{table}
\begin{table}
\caption{{\scriptsize{Low-lying $SU(3)$ irreps (giving hw irrep, nhw irrep, nnhw irrep etc.) $(\la , \mu)$ for $^{82,84,86,88,90,92}$Pd isotopes. In the table results are shown as $\Gamma(\la\;\;\mu)$ where $\Gamma$ is the multiplicity of the $SU(3)$ irrep $(\la\;\;\mu)$. The irreps with $\Gamma=0$ are not shown as they will not exist. Listed only are those $(\la , \mu)$ irreps with $2\la+\mu=\epsilon_H$, $\epsilon_H-3$ and $\epsilon_H-6$. Shown in the table are also number of valence nucleons $m$, isospin $T$ and the $U(10)$ irrep $\l\{f\r\}$. In addition $\l\{f_h\r\}$ is also
given when appropriate. See text for further details.}}}
{\scriptsize{
\begin{tabular}{l}
\hline
$^{82}$Pd $\;:\;\;(m=26,T=5)\;:\;\;\l\{f\r\}=\l\{4^4,2^5\r\},\;
\l\{f_h\r\}=\l\{4,2^5\r\}$ \\  
\verb|  1(10 10)   1(11  8)   3(12  6)   2(13  4)   3(14  2)   1( 8 11)   4( 9  9) | \\
\verb|  7(10  7)  10(11  5)  10(12  3)   6(13  1)   1( 5 14)   4( 6 12)   9( 7 10) | \\
\verb| 21( 8  8)  28( 9  6)  35(10  4)  25(11  2)  12(12  0) | \\
$^{84}$Pd $\;:\;\;(m=28,T=4)\;:\;\;\l\{f\r\}=\l\{4^5,2^4\r\},\;
\l\{f_h\r\}=\l\{4,2^4\r\}$ \\  
\verb|  1(10 10)   1(11  8)   3(12  6)   2(13  4)   3(14  2)   1( 7 13)   2( 8 11) | \\
\verb|  6( 9  9)   9(10  7)  12(11  5)  11(12  3)   7(13  1)   1( 4 16)   2( 5 14) | \\
\verb|  8( 6 12)  14( 7 10)  28( 8  8)  33( 9  6)  40(10  4)  27(11  2)  14(12  0) | \\
$^{86}$Pd $\;:\;\;(m=30,T=3)\;:\;\;\l\{f\r\}=\l\{4^6,2^3\r\},\;
\l\{f_h\r\}=\l\{4,2^3\r\}$ \\    
\verb|  1(12  6)   1(14  2)   1( 8 11)   2( 9  9)   3(10  7)   4(11  5)   4(12  3) | \\
\verb|  2(13  1)   1( 4 16)   2( 5 14)   6( 6 12)   9( 7 10)  16( 8  8)  17( 9  6) | \\
\verb| 20(10  4)  12(11  2)   7(12  0) | \\
$^{88}$Pd $\;:\;\;(m=32,T=2)\;:\;\;\l\{f\r\}=\l\{4^7,2^2\r\},\;
\l\{f_h\r\}=\l\{4,2^2\r\}$ \\   
\verb|  1( 6 12)   1( 7 10)   3( 8  8)   2( 9  6)   3(10  4)   1(11  2)   1(12  0) | \\
\verb|  1( 3 15)   2( 4 13)   5( 5 11)   7( 6  9)  10( 7  7)  10( 8  5)   9( 9  3) | \\
\verb|  5(10  1)   1( 0 18)   1( 1 16)   5( 2 14)   8( 3 12)  17( 4 10)  21( 5  8) | \\
\verb| 29( 6  6)  24( 7  4)  21( 8  2)   5( 9  0) | \\
$^{90}$Pd $\;:\;\;(m=34,T=1)\;:\;\;\l\{f\r\}=\l\{4^8,2\r\},\;
\l\{f_h\r\}=\l\{4,2\r\}$ \\   
\verb|  1( 2 14)   1( 3 12)   3( 4 10)   2( 5  8)   4( 6  6)   2( 7  4)   3( 8  2) | \\
\verb|  1( 1 13)   2( 2 11)   4( 3  9)   5( 4  7)   6( 5  5)   5( 6  3)   3( 7  1) | \\
\verb|  2( 0 12)   4( 1 10)   8( 2  8)   9( 3  6)  11( 4  4)   7( 5  2)   4( 6  0) | \\
$^{92}$Pd $\;:\;\;(m=36,T=0)\;:\;\;\l\{f\r\}=\l\{4^9\r\},\;
\l\{f_h\r\}=\l\{4\r\}$ \\    
\verb|  1( 0 12)   1( 2  8)   1( 3  6)   1( 4  4)   1( 6  0)   1( 3  3)   1( 0  6) | \\
\verb|  1( 1  4)   1( 2  2) | \\ 
\hline
\end{tabular}
}}
\end{table}

\section{Results for $SU(3)$ irreps for Ge to Pd isotopes with proxy-$SU(4)$ symmetry}

Our main purpose in this article is to  provide $SU(3)$ irreps,
good for low-lying states, for Ge, Se, Kr, Sr, Zr, Mo, Ru 
and Pd nuclei. These nuclei
with neutron number $32 \le N \le 46$, are described with $^{56}$Ni core and the valence protons and neutrons occupy
$^1p_{3/2}, ^0f_{5/2}, ^1p_{1/2}$ and $^0g_{9/2}$ orbits. With proxy-$SU(3)$ scheme, the $^0g_{9/2}$ orbit that is pushed down from
the higher shell due to strong spin-orbit force (giving magic number 50) is replaced by proxy $^0f_{7/2}$ orbit. Then we have the proxy $\eta=3$ shell. With this, the spectrum generating algebra is $U(40)$ and there are two ways to obtain
proxy-$SU(3)$ algebra. One is to use $SU(3)$ algebras for protons and neutrons (called I below) and other is to use $SU(3)$ algebra via proxy-$SU(4)$ algebra (called II below). Then we have,
\be
\barr{rcccl}
I & : & U(40) & \supset & \l\{ \l[U_p(20) \supset \l[\supset U(10) \supset SU_p(3)\r] \otimes SU_{S_p}(2)\r] \r\} \\ 
& & & \oplus & \l\{ \l[U_n(20) \supset \l[U(10) \supset SU_n(3)\r] \otimes SU_{S_n}(2)\r] \r\} \\
& & & \supset & \l[SU_{p+n}(3) \supset SO_{p+n}(3)\r] \otimes SU_{p+n:S}(2)\;. \\ 
II  & : & U(40) & \supset & \l[SU_{p+n}(10) \supset SU_{p+n}(3) \supset SO_{p+n}(3) \r] \otimes \l[SU_{p+n}(4) \supset SU_{p+n:S}(2) \otimes SU_{p+n:T}(2)\r]\;.
\earr \label{su4-13}
\ee
The chain I with proton $SU_{p}(3)$ and neutron $SU_n(3)$ algebras is used in \cite{Bona-5} for Kr to Pd isotopes and it is hereafter called pn-proxy-SU(3) algebra. However, as shown in \cite{Ko-psu4} the proxy-$SU(3)$ with proxy-$SU(4)$ algebra given by II is more appropriate for these nuclei as both valence protons and neutrons in these nuclei occupy the same proxy $\eta=3$ shell. With proxy-$SU(4)$, the lowest $SU(4)$ irrep for even-even nuclei with a given isospin $T=|M_T|=|(N-Z)/2|$, describing low-lying levels (or low-lying rotational bands), are given by Eqs. (\ref{su4-1}) and (\ref{su4-2}). Starting with a given nucleus, we have the number $m$ of valence nucleons, isospin $T$ and the corresponding lowest proxy-$SU(4)$ irrep. The proxy-$SU(4)$ irreps give the corresponding $U(10)$ irreps $\l\{f\r\} = \l\{4^a, 3^b, 2^c, 1^d\r\}$. In all our examples $b=0$ and $d=0$. In the situation that $m > 20$, we use the corresponding hole irrep $\l\{f_h\r\}$ and for $a \ne 0$,
\be
\l\{f\r\} = \l\{4^a, 3^b, 2^c, 1^d\r\} \Rightarrow \l\{f_h\r\} =
\l\{4^{10-a-b-c-d},3^d,2^c,1^b\r\}\;.
\label{su4-14}
\ee 
Let us mention that if $a=0$ but $b \ne 0$, $\l\{f_h\r\} = \l\{3^{10-b-c-d}, 2^d,1^c\r\}$. Similarly, for $a=b=0$, $\l\{f_h\r\} = \l\{2^{10-c-d}, 1^d\r\}$ and for $a=b=c=0$, $\l\{f_h\r\} = \l\{
1^{10-d}\r\}$. With these we use
\be
\barr{l}
\l\{f_h\r\} \rightarrow \dis\sum_i x_i (\la_i , \mu_i) \oplus \\
\Rightarrow \l\{f\r\} \rightarrow \dis\sum_i x_i (\mu_i , \la_i) \oplus\;.
\earr \label{su4-15}
\ee
Note that Eq. (\ref{su4-15}) generalizes Eqs. (\ref{su4-3})
and (\ref{su4-12}). Using the method described in Section II-A,
$SU(3)$ irreps are obtained for Ge, Se, Kr, Sr, Zr, Mo, Ru 
and Pd isotopes. The results are given in Tables III-X. In the
tables, for each nucleus $(m,T)$ and $\{f\}$ (also $\{f_h\}$ for $m > 20$) are given along with all $SU(3)$ irreps $(\la , \mu)$
satisfying $2\la+\mu=\epsilon_H$, $\epsilon_H-3$ and $\epsilon_H-6$. They will give not only hw but also nhw and nnhw irreps as discussed below. It is important to mention that spin $S=0$ for all the irreps listed in the Tables III-X. 

\subsection{Results for Ge isotopes}

In Table III, $SU(3)$ irreps for $^{64,66,68,70,72,74,76}$Ge are
listed. Ignoring the $(\la , \mu)$ irreps with $\la$ odd or $\mu$ odd, as seen from Table III, the hw, nhw and nnhw irreps for $^{64}$Ge are $[(16,4), (12,6)^2 , (14,2)^3]$. Similarly, for $^{66}$Ge they are $[(20,2), (14,8) , (16,4)^6]$, for $^{68}$Ge they are $[(18,6), (20,2)^2 , (14,8)^7]$, for $^{70}$Ge they are $[(18,6), (20,2)^2 , (12,12)]$, for $^{72}$Ge they are $[(20,2), (12,12) , (14,8)^6]$, for $^{74}$Ge they are $[(14,8), (16,4)^2 , (18,0)]$ and finally for $^{76}$Ge they are $[(10,10), (12,6)^3 , (14,2)^3]$. Thus, except for hw irrep, in general the nhw irrep and nnhw irrep carry multiplicities.
Although the hw and nhw irreps are not listed for Ge isotopes in
\cite{Bona-5}, the above results are useful as Ge isotopes are considered for example in \cite{Bona-7} in the context of triaxiality in nuclei. 

\subsection{Results for Se isotopes}

In Table IV, $SU(3)$ irreps for $^{70,72,74,76,78,80,82}$Se are
listed. Ignoring the $(\la , \mu)$ irreps with $\la$ odd or $\mu$ odd, as seen from Table IV, the hw, nhw and nnhw irreps for $^{70}$Se are $[(22,4), (24,0), (16,10)]$. Similarly, for $^{72}$Se they are $[(22,4), (24,0) , (16,10)^3]$, for $^{74}$Se they are $[(24,0), (16,10), (18,6)^7]$, for $^{76}$Se they are $[(18,6), (20,2), (12,12)^3]$, for $^{78}$Se they are $[(14,8), (16,4)^2, (18,0)]$, for $^{80}$Se they are $[(12,6), (14,2), (6,12)]$ and finally for $^{82}$Se they are $[(12,0), (6,6), (8,2)^3]$. Thus, except for hw irrep, in general the nhw irrep and nnhw irrep carry multiplicities.
Although the hw and nhw irreps are not listed for Se isotopes in
\cite{Bona-5}, the above results are useful as Se isotopes are considered for example in \cite{Bona-7} in the context of triaxiality in nuclei. 
   
\subsection{Results for Kr isotopes}

In Table V, $SU(3)$ irreps for $^{72,74,76,78,80,82}$Kr are
listed. Ignoring the $(\la , \mu)$ irreps with $\la$ odd or $\mu$ odd, as seen from Table V, the hw, nhw and nnhw irreps for $^{72}$Kr are $[(20,8), (22,4), (24,0)]$. Similarly, for $^{74}$Kr they are $[(20,8), (22,4)^2, (24,0)]$, for $^{76}$Kr they are $[(22,4), (24,0), (14,14)]$, for $^{78}$Kr they are $[(16,10), (18,6)^3, (20,2)^3]$, for $^{80}$Kr they are $[(12,12), (14,8)^4, (16,4)^6]$, and finally for $^{82}$Kr they are $[(10,10), (12,6)^3, (14,2)^3]$. Thus, except for hw irrep, in general the nhw irrep and nnhw irrep carry multiplicities.
The hw and nhw irreps given here, except for $^{76}$Kr, are same as those listed for the corresponding Kr isotopes in \cite{Bona-5} where the pn-proxy-SU(3) algebra is used. For $^{76}$Kr, the nhw irrep is $(24,0)$ with proxy-$SU(4)$ and the nnhw irrep is $(14,14)$. The irrep $(14,14)$ is the nhw irrep with pn-proxy-SU(3) algebra as given in \cite{Bona-5}. 
Also, in \cite{Bona-5} multiplicities of nhw irreps are ignored. The multiplicities are important as they provide a weight factor when the $(\beta,\gamma)$ values for hw irrep and nhw irrep are combined. Appendix A gives the formulas for the $(\beta,\gamma)$
parameters. 

\subsection{Results for Sr isotopes}

In Table VI, $SU(3)$ irreps for $^{74,76,78,80,82,84}$Sr are
listed. Ignoring the $(\la , \mu)$ irreps with $\la$ odd or $\mu$ odd, as seen from Table VI, the hw, nhw and nnhw irreps for $^{74}$Sr are $[(20,8), (22,4)^2, (24,0)]$. Similarly, for $^{76}$Sr they are $[(20,8), (22,4), (24,0)]$, for $^{78}$Sr they are $[(22,4), (24,0), (14,14)]$, for $^{80}$Sr they are $[(16,10), (18,6)^3, (20,2)^3]$, for $^{82}$Sr they are $[(12,12), (14,8)^4, (16,4)^6]$, and finally for $^{84}$Sr they are $[(10,10), (12,6)^3, (14,2)^3]$. Thus, except for hw irrep, in general the nhw irrep and nnhw irrep carry multiplicities.
The hw and nhw irreps given here, except for $^{78}$Sr, are same as those listed for the corresponding Sr isotopes in \cite{Bona-5} where the pn-proxy-SU(3) algebra is used. For $^{78}$Sr, the nhw irrep is $(24,0)$ with proxy-$SU(4)$ and the nnhw irrep is $(14,14)$. The irrep $(14,14)$ is the nhw irrep with pn-proxy-SU(3) algebra as given in \cite{Bona-5}. 
Also, in \cite{Bona-5} multiplicities of nhw irreps are ignored. As mentioned before, multiplicities are important as they provide a weight factor when the $(\beta,\gamma)$ values for hw irrep and nhw irrep are combined.

\subsection{Results for Zr isotopes}

In Table VII, $SU(3)$ irreps for $^{76,78,80,82,84,86}$Zr are
listed. Ignoring the $(\la , \mu)$ irreps with $\la$ odd or $\mu$ odd, as seen from Table VII, the hw, nhw and nnhw irreps for $^{76}$Zr are $[(22,4), (24,0), (14,14)]$. Similarly, for $^{78}$Zr they are $[(22,4), (24,0), (14,14)]$, for $^{80}$Zr they are $[(24,0), (16,10), (18,6)^3]$, for $^{82}$Zr they are $[(18,6), (20,2), (10,16)]$, for $^{84}$Zr they are $[(14,8), (16,4)^2, (18,0)]$, and finally for $^{86}$Zr they are $[(12,6), (14,2), (4,16)]$. Thus, except for hw irrep, in general the nhw irrep and nnhw irrep carry multiplicities.
The hw irreps given here are same as those listed for the corresponding Zr isotopes in \cite{Bona-5}. However, for the nhw irreps there are differences. In \cite{Bona-5}, where the 
pn-proxy-SU(3) algebra is used,
the nhw irreps for $^{76}$Zr, $^{78}$Zr and $^{80}$Zr are given to be $(16,10)$, $(16,10)$ and $(8,20)$ respectively. With proxy-$SU(4)$, as seen from Table VII, these are much lower irreps compared to even the nnhw irreps. Similarly, the nhw irreps for $^{82}$Zr, $^{84}$Zr and $^{86}$Zr, with pn-proxy-SU(3) algebra, are $(10,16)$, $(6,18)$ and $(4,16)$ respectively (D. Bonatsos, private communication). For $^{82}$Zr and $^{86}$Zr these nhw irreps are same as the nnhw irreps with proxy-$SU(4)$ symmetry as can be seen from Table VII. However, for $^{84}$Kr the nhw irrep with pn-proxy-SU(3) algebra is $(6,18)$ and this is much lower than the nnhw irrep with proxy-$SU(4)$ algebra. Also,
just as before, in \cite{Bona-5} multiplicities of nhw irreps are ignored. 

\subsection{Results for Mo isotopes}

In Table VIII, $SU(3)$ irreps for $^{78,80,82,84,86,88}$Mo are
listed. Ignoring the $(\la , \mu)$ irreps with $\la$ odd or $\mu$ odd, as seen from Table VIII, the hw, nhw and nnhw irreps for $^{78}$Mo are $[(16,10), (18,6)^3, (20,2)^3]$. Similarly, for $^{80}$Mo they are $[(16,10), (18,6)^3, (20,2)^3]$, for $^{82}$Mo they are $[(18,6), (20,2), (10,16)]$, for $^{84}$Mo they are $[(12,12), (14,8), (16,4)]$, for $^{86}$Mo they are $[(8,14), (10,10)^3, (12,6)^4]$, and finally for $^{88}$Mo they are $[(6,12), (8,8)^3, (10,4)^3]$. Thus, except for hw irrep, in general the nhw irrep and nnhw irrep carry multiplicities.
The hw and nhw irreps given here, except for $^{82}$Mo, are same as those listed for the corresponding Mo isotopes in \cite{Bona-5} where the pn-proxy-SU(3) algebra is used. For $^{82}$Mo, the nhw irrep is $(20,2)$ with proxy-$SU(4)$ and the nnhw irrep is $(10,16)$. The irrep $(10,16)$ is the nhw irrep with pn-proxy-SU(3) algebra as given in \cite{Bona-5}. Also, in \cite{Bona-5} multiplicities of nhw irreps are ignored.   

\subsection{Results for Ru isotopes}

In Table IX, $SU(3)$ irreps for $^{80,82,84,86,88,90}$Ru are
listed. Ignoring the $(\la , \mu)$ irreps with $\la$ odd or $\mu$ odd, as seen from Table IX, the hw, nhw and nnhw irreps for $^{80}$Ru are $[(12,12), (14,8)^4, (16,4)^6]$. Similarly, for $^{82}$Ru they are $[(12,12), (14,8)^4, (16,4)^6]$, for $^{84}$Ru they are $[(14,8), (16,4)^2, (18,0)]$, for $^{86}$Ru they are $[(8,14), (10,10)^3, (12,6)^4]$, for $^{88}$Ru they are $[(4,16), (6,12)^2, (8,8)^3]$, and finally for $^{90}$Ru they are $[(2,14), (4,10)^3, (6,6)^4]$. Thus, except for hw irrep, in general the nhw irrep and nnhw irrep carry multiplicities.
The hw and nhw irreps given here, except for $^{84}$Ru, are same as those listed for the corresponding Ru isotopes in \cite{Bona-5} where the pn-proxy-SU(3) algebra is used. For $^{84}$Ru, the nhw irrep is $(16,4)$ with proxy-$SU(4)$ while in \cite{Bona-5} the nhw irrep is given to be $(6,18)$ and it is much lower than the nnhw irrep with proxy-$SU(4)$. 
Also, in \cite{Bona-5} multiplicities of nhw irreps are ignored.   
  
\subsection{Results for Pd isotopes}

In Table X, $SU(3)$ irreps for $^{82,84,86,88,90,92}$Pd are
listed. Ignoring the $(\la , \mu)$ irreps with $\la$ odd or $\mu$ odd, as seen from Table X, the hw, nhw and nnhw irreps for $^{82}$Pd are $[(10,10), (12,6)^3, (14,2)^3]$. Similarly, for $^{84}$Pd they are $[(10,10), (12,6)^3, (14,2)^3]$, for $^{86}$Pd they are $[(12,6), (14,2), (4,16)]$, for $^{88}$Pd they are $[(6,12), (8,8)^3, (10,4)^3]$, for $^{90}$Pd they are $[(2,14), (4,10)^3, (6,6)^4]$, and finally for $^{92}$Pd they are $[(0,12), (2,8), (4,4)]$. Thus, except for hw irrep, in general the nhw irrep and nnhw irrep carry multiplicities.
The hw and nhw irreps given here, except for $^{86}$Pd, are same as those listed for the corresponding Pd isotopes in \cite{Bona-5} where the pn-proxy-SU(3) algebra is used. For $^{86}$Pd, with proxy-$SU(4)$ the nnhw irrep is $(4,16)$. This irrep is the nhw irrep with pn-proxy-SU(3) algebra as given in \cite{Bona-5}. Also, in \cite{Bona-5} multiplicities of nhw irreps are ignored.

\section{Conclusions}

Recently it is shown that for nuclei with $32 \le \mbox{Z,N} \le 46$ proxy-$SU(4)$ symmetry is important. Therefore, in the analysis of these nuclei for prolate-oblate transition, shape 
coexistence and triaxiality one needs proxy-$SU(3)$ algebra along with the spin-isospin proxy-$SU(4)$ algebra. As these nuclei occupy proxy $\eta=3$ shell, the appropriate algebra  is then,  
$$
U(40) \supset \l[U(10) \supset SU(3) \supset SO(3)\r] \otimes \l[SU(4) \supset SU_S(2) \otimes SU_T(2)\r]\;.
$$
Following this, presented in this paper are the $SU(3)$ irreps $(\lambda ,\mu)$ with $2\lambda + \mu =2\lambda_H + \mu_H-3r$, $r=0,1,2$ for various isotopes of Ge, Se, Kr, Sr, Zr, Mo, Ru and Pd assuming good proxy-$SU(4)$ symmetry. Here $(\la_H , \mu_H)$ is the hw $SU(3)$ irrep. Results are presented
in Section III in Tables III-X . These results are generated using a simple method described in Section II and as a test used are two columned $U(10)$ irreps and they are presented in Tables I and II. Results in all the tables are compared with those given
by Bonatsos et al in Refs. \cite{Bona-5,Bona-7} and the differences are discussed in detail. The tabulations for proxy-$SU(3)$ irreps provided in this paper, for Ge, Se, Kr, Sr, Zr, Mo, Ru and Pd isotopes with good proxy-$SU(4)$ symmetry are useful in further investigation of triaxial shapes in these nuclei and this application will be discussed in a future publication.  

\acknowledgments

This work is a result of on-line participation in the Sofia meetings held in 2021, 2023 and 2025. Thanks are due to N. Minkov for these meetings.

\renewcommand{\theequation}{A-\arabic{equation}}
    \setcounter{equation}{0}
    \section*{APPENDIX A}

Given a $SU(3)$ irrep $(\la , \mu)$, the corresponding Bohr-Mottelson quadrupole shape parameters $(\beta,\gamma)$ are given by the following formulas \cite{Bona-5,jpd-bg},
\be
\barr{rcl}
\beta^2 & = & \dis\frac{4\pi}{5\l(A\overline{r^2}\r)^2}\,\l(\la^2 + \mu^2 + 
\la \mu + 3\la + 3\mu +3\r)\;,\\
\gamma & = & \arctan \l[\dis\frac{\dis\sqrt{3}(\mu +1)}{(2\la +\mu +3)}\r]\;.
\earr \label{eq.BM-6}
\ee 
Here, $A$ is nucleon number and $\overline{r^2}$ is the dimensionless mean square
radius. It is well known that $(\overline{r^2})^{1/2} = r_{r.m.s} = r_0 A^{1/6}$ where $r_0 = 0.87$.

\end{document}